\documentclass[aps,prl,twocolumn,floatfix,superscriptaddress,showpacs,longbibliography]{revtex4-1}
\usepackage{graphicx}
\usepackage{amsmath}
\usepackage{amssymb}
\usepackage{amsfonts}
\usepackage{bm}        
\usepackage{dsfont}
\usepackage[mathscr]{eucal}
\usepackage[colorlinks, linkcolor=OrangeRed,citecolor=RoyalBlue,urlcolor=NavyBlue]{hyperref}
\usepackage[normalem]{ulem}
\usepackage[all]{hypcap}
\usepackage[dvipsnames,usenames,table]{xcolor}

\newcommand{\vk}{\mathbf{k}}

\newcommand{\eq}[1]{\begin{align}#1\end{align}}

\newcommand{\nn}{\nonumber}
\newcommand{\pr}{\mathbf{r}}

\newcommand{\mcl}[1]{\mathcal{#1}}
\newcommand{\mrm}[1]{\mathrm{#1}}

\begin{document}

\title{ Emergent Weyl nodes and Fermi arcs in a  Floquet Weyl semimetal
}

\author{Leda Bucciantini}
\affiliation{Max-Planck-Institut f{\"u}r Physik komplexer Systeme, N{\"o}thnitzer Stra{\ss}e 38, 01187 Dresden, Germany}
\affiliation{Max-Planck-Institut f{\"u}r Chemische Physik fester Stoffe, N{\"o}thnitzer Stra{\ss}e 40, 01187 Dresden, Germany}

\author{Sthitadhi Roy}
\affiliation{Max-Planck-Institut f{\"u}r Physik komplexer Systeme, N{\"o}thnitzer Stra{\ss}e 38, 01187 Dresden, Germany}

\author{Sota Kitamura}
\affiliation{Department of Physics, University of Tokyo, Hongo, Tokyo 113-0033, Japan}

\author{Takashi Oka} 
\affiliation{Max-Planck-Institut f{\"u}r Physik komplexer Systeme, N{\"o}thnitzer Stra{\ss}e 38, 01187 Dresden, Germany}
\affiliation{Max-Planck-Institut f{\"u}r Chemische Physik fester Stoffe, N{\"o}thnitzer Stra{\ss}e 40, 01187 Dresden, Germany}

\date{\today}
\begin{abstract}
When a Dirac semimetal is subject to a circularly polarized laser, it is predicted that the Dirac cone splits into 
two Weyl nodes and a nonequilibrium transient state called the Floquet Weyl semimetal is realized. 
We focus on the previously unexplored low frequency regime, where the upper and lower Dirac bands 
resonantly couples with each other through multi-photon processes, which is a realistic situation in solid state 
ultrafast pump-probe experiments. 
We find a series of new Weyl nodes emerging in pairs when the Floquet replica bands hybridize with each other. 
The nature of the Floquet Weyl semimetal with regard to the number, locations, and  monopole charges of these Weyl nodes is highly tunable with the amplitude and frequency of the light. We derive an effective low energy theory using Brillouin-Wigner expansion and further regularize the theory on a cubic lattice. The monopole charges obtained from the low-energy Hamiltonian can be reconciled with the number of Fermi arcs on the lattice which we find numerically.
\end{abstract}

\maketitle

\paragraph{Introduction:}
Weyl semimetals (WSM) and Dirac semimetals (DSM) have emerged as one of the most exciting new class of three-dimensional topological materials \cite{Murakami07,PhysRevLett.107.127205, Hosur:2013eb, Turner:2013tf, Vafek:2014hl,jia2016weyl} with a gapless and linearly dispersing bulk spectrum allowing for a realization of Weyl fermions. Since the Weyl nodes in a lattice always occur in pairs of opposite chiralities \cite{NielNino81c}, they act as  monopoles and anti-monopoles of Berry flux. Consequently, WSMs have topological surface states whose Fermi surfaces originate and terminate at Weyl nodes of opposite chiralities leading to open Fermi arcs \cite{Wan:2011hi, Hosur:2012hr, Ojanen:2013jn, 2014PhRvB..89w5315O,  Potter:2014cc,SIN15, XAB15, Huang2015}. The non-trivial topology of WSMs leads to various exotic electromagnetic responses like the condensed matter realization of the chiral anomaly \cite{NielNino83, PhysRevD.86.045001,Zyuzin2012kl,PhysRevB.88.104412,Landsteiner:2014fw, 2016PhRvB..93g5114B,2016PhRvB..94p1107R}, chiral magnetic effect, \cite{2008PhRvD..78g4033F, 2012PhRvL.109r1602S}  and  negative magnetoresistance \cite{PhysRevB.88.104412, 2015PhRvB..91x5157B,2016NatCo...711615A}. Although WSM materials have been discovered recently,\cite{SIN15,Lv2015,XAB15,LvXu2015,YLS15,Huang2015} it is extremely desirable to posses the capability of tuning their properties with regard to the number and nature of Weyl nodes. 
It is known that WSM can be generated out of a DSM when time-reversal (TRS) and/or inversion symmetries are broken \cite{PhysRevLett.107.127205,Zyuzin:2012ca,HB12,Weng2015}. While the former separates  the Weyl nodes in momentum, the latter separates them in energy.

Recently, time periodic modulations of topologically trivial systems, often realized via light-matter interaction, have emerged as an interesting way of obtaining topological phases, often richer than their static counterparts \cite{OA2009,IT2010,lindner2011floquet,GFAA2011,KOBFD2011,LBDRG2013,cayssol2013floquet,DGP2013,KP2013,UPFB2014,DOM2015,d2015dynamical,PhysRevB.93.184306} 
and effects of interaction and disorder have also been explored \cite{Rudner13,Titum2015, TBRRL2016, Khemani2016,Sreejith2016,roy2016disordered}.  Such protocols have also been complemented with their experimental realizations.\cite{wang2013observation,mahmood2016selective}

In the same spirit, one expects to generate a three-dimensional topological WSM from its trivial parent, a Dirac semimetal (DSM) by subjecting it to time-periodic fields \cite{ 2014EL....10517004W, 2016PhRvB..93o5107E,PhysRevLett.116.026805,2016arXiv160403399H,2016PhRvL.117h7402Y,zhang2016theory}.
An appealing way of breaking TRS in the context of solid state experiments is to subject DSM materials like Na$_3$Bi and Cd$_3$As$_2$\cite{PhysRevB.92.075115,liu2014stable,PhysRevB.91.241114,neupane2014observation,liang2015ultrahigh} to circularly polarized laser (CPL). 
Such a system is described by a Hamiltonian periodic in time and hence can be studied using Floquet theory \cite{PhysRevA.7.2203, PhysRev.138.B979, grifoni1998driven}.
Analysis of the Floquet quasienergy spectrum reveals a new Floquet WSM phase born out of the DSM, in which the number, location, and nature of the Weyl nodes are tunable with the amplitude and frequency of CPL.
Previously, such a system has been studied within the framework of high-frequency Floquet-Magnus expansion. \cite{effMag1, effMag2}

%
\begin{figure*}[t]
\begin{center}
\includegraphics[width=\linewidth]{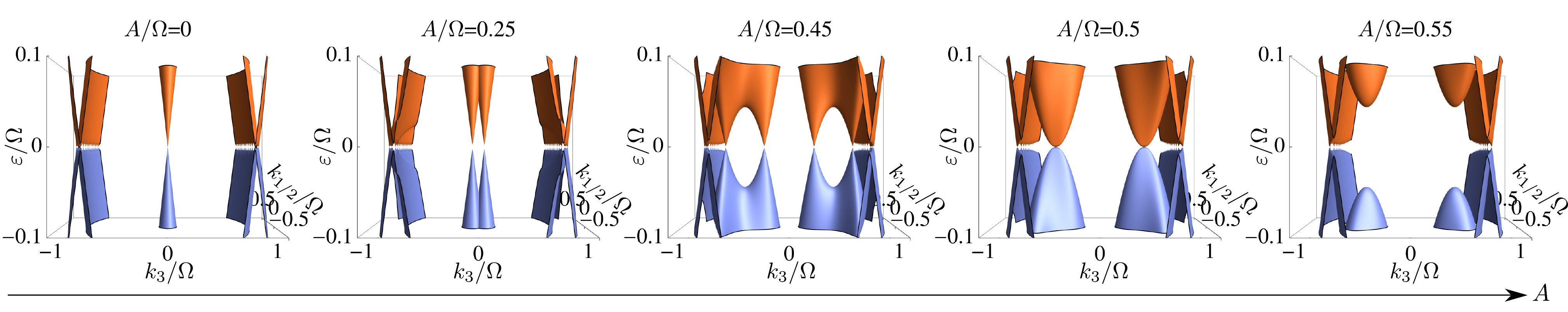}
\caption{Evolution of the  Floquet quasienergy spectrum on increasing laser amplitude $A$. 
Weyl points generated from the $n=0$ and $n=\pm1$ sectors at finite $A$ meet and annihilate at $A/\Omega=0.5$.
  }
\label{fig:weyltraj}
\end{center}
\end{figure*}

The focus of this work, however, is the much richer and experimentally relevant regime, namely the situation where the frequency of the CPL is much less than or comparable to the bandwidth of the parent DSM.
In this regime, Floquet replica, i.e. photon dressed states, will cross, hybridize with and repel each other.  As a consequence, besides the two Weyl nodes born out of the original Dirac cone, we find a series of infinite number of Weyl nodes emerging from the Floquet replicas. They have nontrivial monopole charges, and as the CPL amplitude is increased, they move and pairwise annihilate as they approach those with opposite monopole numbers (see Fig.~\ref{fig:weyltraj} for example).

We derive effective low-energy Hamiltonians for these new Weyl nodes using Brillouin-Wigner expansion \cite{hubac2000use, young1957continued, 2016PhRvB..93n4307M} and deduce their monopole charges. 
Similar to the static case, Floquet Fermi arcs are generated between the Weyl nodes and their degeneracy is related to the monopole number of the Weyl nodes. These results are verified by regularizing our theory on a lattice, and numerically computing the number of Fermi arcs in a system with open boundaries.
Our findings can be experimentally realized using time resolved ARPES or ultrafast pump-probe measurements which we will explain in the end.

\paragraph{Spectrum of the Floquet WSM:}
Floquet theory reduces the solution of the Schr\"odinger equation for a time periodic Hamiltonian $\mcl{H}(t+T)=\mcl{H}(t)$ with $T=2\pi/\Omega$, to an eigenvalue problem for a time-independent, but infinite dimensional, Hamiltonian (for a review see \cite{EckardtRMP16}). The infinite dimensional Floquet Hamiltonian, $\mcl{H}^F$, has blocks  of the general form 
\eq{
\mcl{H}^F_{m,n}=\frac{1}{T}\int_{-T/2}^{T/2}dt~e^{i(m-n)\Omega t}\mcl{H}(t) - n\Omega\delta_{m,n},
\label{eq:hmn}
}
 where $n,m\in\mathds{Z}$.
The diagonal block $\mcl{H}^{F}_{n,n}$ corresponding to the $n$-photon sector, is equal to the time averaged Hamiltonian over a period $\mcl{H}^F_0$,  shifted in energy by $n\Omega$. The off-diagonal blocks  $\mcl{H}^{F}_{m,n}=\mcl{H}^{F}_{m-n}$, with $m\neq n$, correspond to transitions between these sectors via absorption or emission of $m-n$ photons.

We start with a continuum low energy description of the DSM  described by the Hamiltonian $\mcl{H}_{\text{DSM}} = \gamma^0(Mc^2+c\hbar\sum_\vk \bm{\gamma}\cdot\vk)$, where $M$ is the mass and $c$ the Fermi velocity  of the DSM .
The CPL propagating along the $\hat{z}$, described by the gauge field $\mathbf{A}(\pr,t)$, is assumed to have a wavelength much larger than the width of the DSM along $\hat{z}$. 
Further, since the magnetic field of the CPL is negligible compared to the electric field, the $\pr$-dependence can be neglected, which reduces the form of the gauge field to $\mathbf{A}(t)=A\{\cos(\Omega t),\sin(\Omega t),0\}$, where $E=A\Omega$ is the amplitude of the electric field and $\Omega$ is the frequency of the CPL. 
The continuum time-dependent Hamiltonian describing the DSM subjected to CPL can then be obtained via minimal coupling between $\mcl{H}_{\text{DSM}}$ and $\mathbf{A}(t)$ as
\eq{
\mcl{H}(t)=\gamma^0\left[Mc^2+\hbar c\sum_\vk \bm{\gamma}\cdot\left(\vk-\frac{e}{\hbar}\mathbf{A}(t)\right)\right].
\label{eq:hminimal}
}
In the  following, we use natural units with $\hbar=1=e=c$ and also set the mass  $M$ of the DSM to zero.
The Floquet Hamiltonian corresponding to $\mcl{H}(t)$ \eqref{eq:hminimal} can be obtained using Eq.~\eqref{eq:hmn} as
\eq{
\mcl{H}^F_0=\gamma^0\sum_\vk\bm{\gamma}\cdot\vk;~~\mcl{H}^F_{\pm1}=-A\gamma^0\gamma^\pm,
\label{eq:hfloq}
}
and $\mcl{H}^F_{n}=0~\forall\vert n\vert>1$, where $\gamma^\pm = (\gamma^1\pm \imath\gamma^2)/2$.
To obtain the Floquet quasienergy spectrum, i.e., the eigenvalues $\varepsilon$ of the Floquet Hamiltonian, $\mcl{H}^F$ is truncated to include a finite number of photon processes and numerically diagonalized.
%

\begin{figure}[t]
\includegraphics[width=\columnwidth]{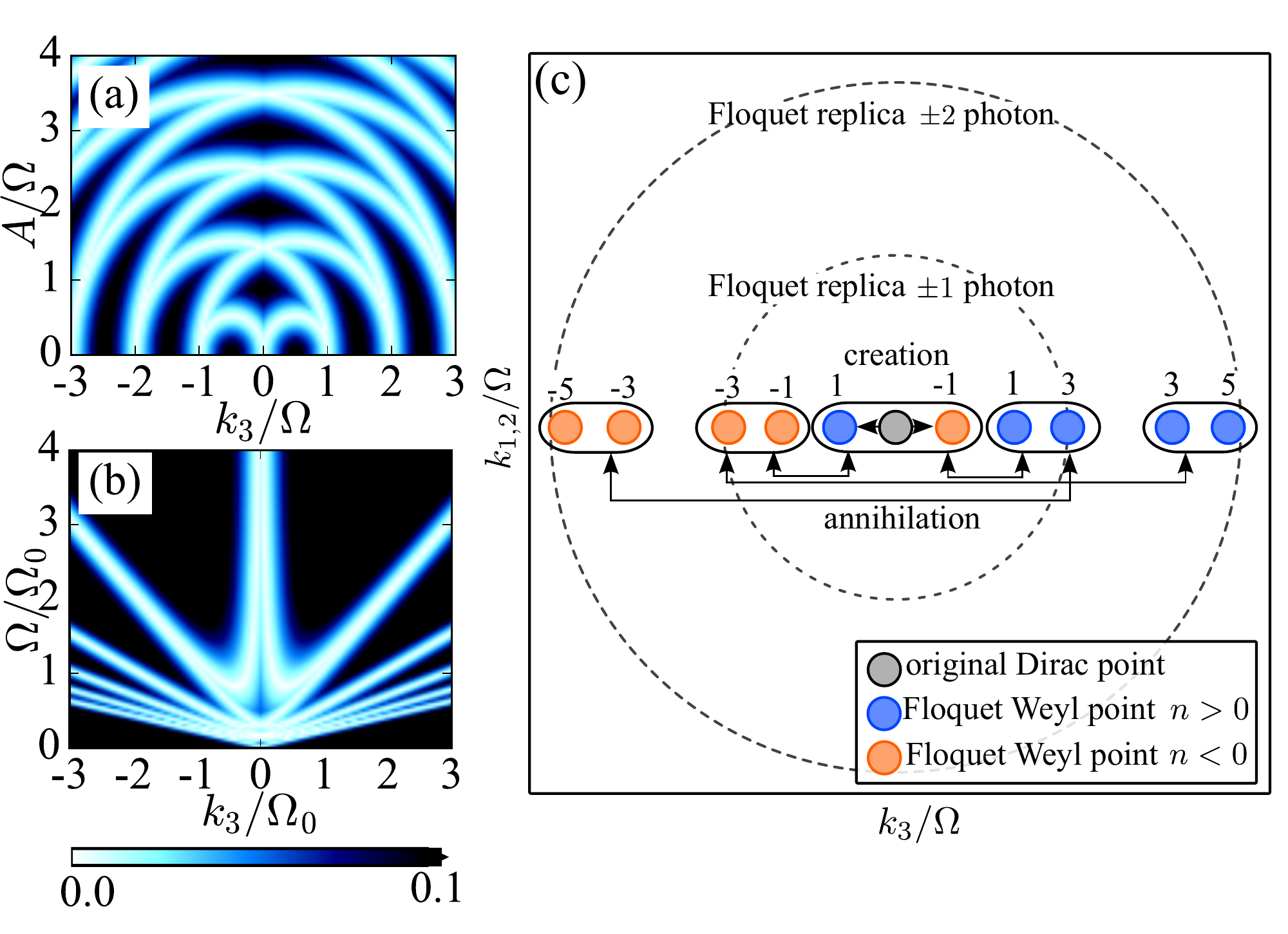}
\caption{The number and location of Weyl points along $k_3$ calculated from the continuum model as function of (a) $A$ and (b) $\Omega$ at a fixed $A=\Omega_0/4$, where $\Omega_0$ is an arbitrarily chosen frequency. The Weyl points are located by plotting the quasienergy spectrum of the band closest to (but above) $\varepsilon=0$ as a color/density plot, where the white lines denote the locus of $\varepsilon=0$. Note that, on decreasing $\Omega$, the spectrum has increasing number of Weyl points close to $k_3=0$ originating from higher order resonances. (c) Schematic plot showing the creation and annihilation of emergent Weyl nodes. The dashed circles show the Floquet replica bands that cross the $\varepsilon=0$ surface at $\vert\vk\vert=n\Omega$. The number above each Weyl point indicates its monopole charge.}
\label{fig:annihilation}
\end{figure}

The quasienergy spectrum reveals an exotic Floquet WSM phase, where the number and locations of the Weyl nodes can be tuned via $A$ and $\Omega$ as shown in Fig.~\ref{fig:weyltraj}. 
In the absence of CPL ($A=0$), the Floquet quasienergy spectrum consists of a doubly degenerate Dirac cone at $\vk=0$ and quasienergy $\varepsilon=0$.
Additionally, the gap between the Floquet replicas of the spectrum from the $\pm n$-photon sectors closes on the hypersphere $\vert\vk\vert=n\Omega$ at $\varepsilon=0$. 
A finite $A$ causes  the  Dirac cones at $\vk=0$ to hybridize with each other, and similarly the doubly-degenerate gapless hyperspheres   at $\vert\vk\vert=n\Omega$. Each of these split into two Weyl nodes, lying on the  $k_3$-axis.
On increasing $A$, the  Weyl nodes move along $k_3$, such that they eventually  merge and annihilate each other, resulting in a gapped spectrum.
The global picture is shown in Fig.~\ref{fig:annihilation}(a)-(b) via the trajectories of the Weyl nodes.

It is important to note that, although the continuum theory  is scale-invariant because of its linear dispersion, leading to any $\Omega$ being resonant, it correctly describes the physics of a lattice regularized theory only close to $\vk=0$ within the region where the linear approximation for the DSM spectrum $\sin\vert\vk\vert\sim\vert \vk\vert$ holds.
The spectrum of a DSM on a lattice would have a bounded spectrum with a finite bandwidth.
The resonant regime in this case, unlike the high-frequency limit, allows for hybridization between the $n=\pm1$ Floquet replicas \cite{PhysRevA.91.043625} resulting in additional Weyl nodes as shown for small $\Omega/\Omega_0$ values in Fig.~\ref{fig:annihilation}(b).
It can be estimated that the continuum theory correctly captures the resonances between upto $\pm n$-photon sectors as long as $\Omega$ is small enough such that $\sin(n\Omega)\sim n\Omega$.
Having established that the resonant limit of the Floquet WSM indeed leads to new Weyl nodes from resonances between higher photon sectors, we now obtain effective low-energy Hamiltonians for these new Weyl nodes originating from the hybridization of $n=\pm 1$ Floquet replicas, i.e. near $k_3\simeq \pm \Omega$, to leading order in $A$.
%

\paragraph{Effective theory for the emergent Weyl nodes:}
The Weyl nodes that are created in the Floquet replica bands have nontrivial monopole numbers
and we can construct their effective theories using the Brillouin-Wigner (BW) expansion, where the details can be found in the supplementary material \cite{supp}. 
We note that this method is equivalent to the Green's Function Decimation (GFD) technique \cite{PhysRevB.37.5723,grosso2014solid} which has also been applied to study Floquet states in graphene~\cite{PhysRevA.91.043625}.

\begin{figure}[t]
\includegraphics[width=\columnwidth]{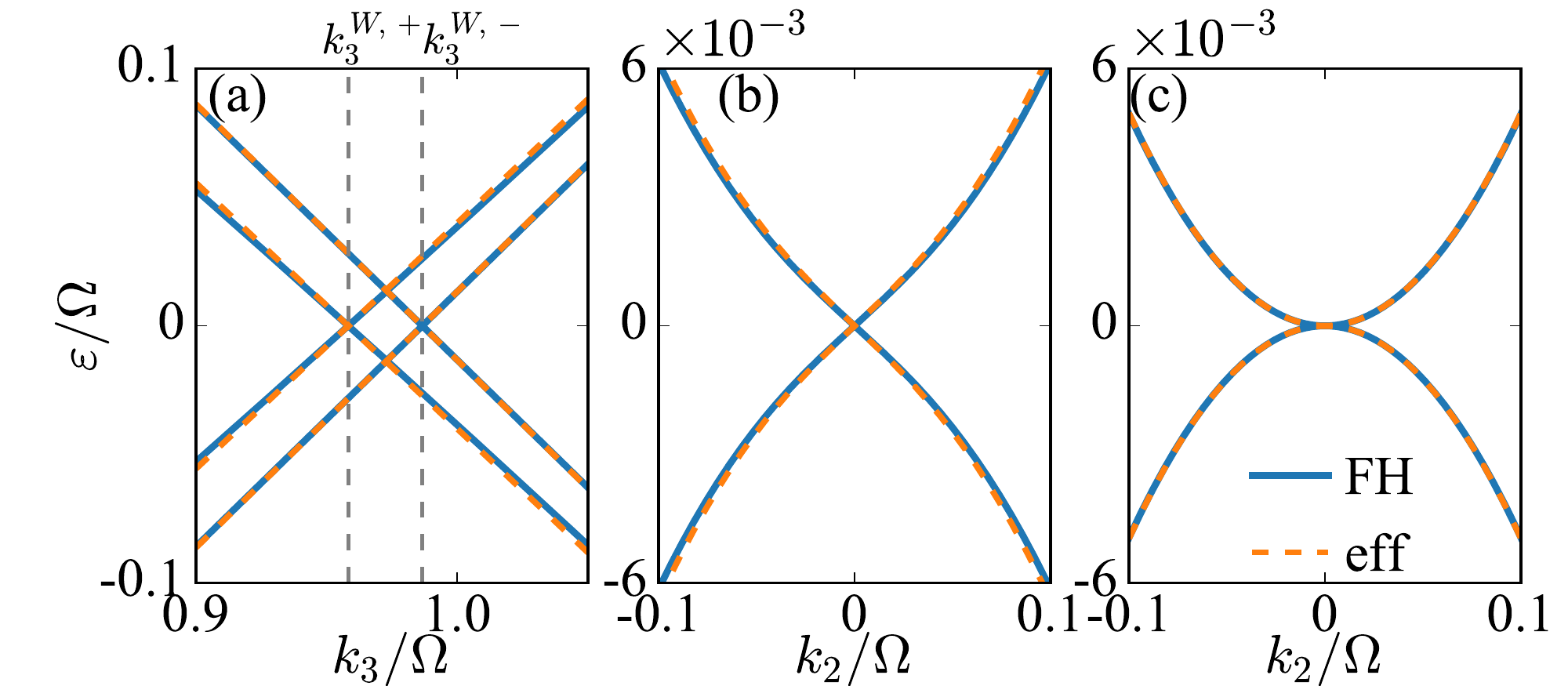}
\caption{Floquet quasienergy spectrum near the Weyl nodes for $A=\Omega/5$.  The spectrum from the Floquet Hamiltonian (FH) \eqref{eq:hfloq} and effective theory obtained from \eqref{eq:bw-weyl} are compared. (a) A slice along $k_3$ with  $k_{1,2}=0$. 
A slice along $k_2$ for (b) $k_3=k_3^{W,+}$ and (c) $k_3=k_3^{W,-}$ at $k_1=0$.
}
\label{fig:comp}
\end{figure}

Here, we demonstrate this for the first Weyl node pairs created by hybridization of the $n=\pm 1$ Floquet replicas that are resonant at $k_3\simeq  \Omega$ and $k_{1,2}=0$. 
The derivation is done by projecting the infinite-dimensional Floquet Hamiltonian \eqref{eq:hfloq} onto the relevant photon sectors that participate in the resonance. Their effective coupling is derived up to leading orders in $A$ by projecting out other photon sectors. 
Given the eigenvalue problem
$\sum_n(\mathcal{H}^{F}_{m,n}-\delta_{m,n}m\Omega)|\Psi_n\rangle=\varepsilon|\Psi_m\rangle$, we aim to reduce it to
an eigenvalue problem 
$\sum_n'\mathcal{H}^{\text{BW}}_{m,n}{P}|\Psi_n\rangle=\varepsilon{P}|\Psi_m\rangle$ in a smaller Hilbert space.  $P$ is the projection operator to the $n=\pm1$ subspace, $P_{m,n}=\delta_{m,1}\delta_{n,1}+\delta_{m,-1}\delta_{n,-1}$ and the sum $\sum'$ is restricted to this space. 
The BW Hamiltonian depends on the exact eigenvalue $\varepsilon$ and up to $A^2$ it is expressed as
\begin{align}
\mathcal{H}_{s,s}^{\text{BW}}(\varepsilon)=&\mathcal{H}^F_{0}-s\Omega+\mathcal{H}^F_{+s}(\varepsilon-\mathcal{H}^F_{0})^{-1}\mathcal{H}^F_{-s}\notag\\
&+\mathcal{H}^F_{-s}(\varepsilon+2s\Omega-\mathcal{H}^F_{0})^{-1}\mathcal{H}^F_{-s};\notag\\
\mathcal{H}_{s,-s}^{\text{BW}}(\varepsilon)=&\mathcal{H}^F_{+s}(\varepsilon-\mathcal{H}^F_{0})^{-1}\mathcal{H}^F_{+s},
\end{align}
with $s=\pm1$. By using the basis diagonal in both $\mathcal{H}^F_0$ and $\gamma^5$, one can further project out the irrelevant states far from $\varepsilon=0$, and replace $\varepsilon$ with an explicit form obtained by an expansion in $A$.
The effective Hamiltonian so obtained has the same block diagonal structure as obtained from GFD and the two blocks are given by
\eq{
\mcl{H}_{\mrm{eff}}^{W,\pm}=&\left(\vert\vk\vert-\Omega+A^{2}\frac{\vert\vk\vert^2+k_{3}^{2}\pm\Omega k_{3}}{\vert\vk\vert(4\vert\vk\vert^{2}-\Omega^{2})}\right)\sigma^3\notag\\
&-\frac{A^{2}(\vert\vk\vert+k_{3})^{\pm1}}{2\vert\vk\vert\Omega(2\vert\vk\vert-\Omega)}(k_+^{2\mp1}\sigma^++\text{h.c.}),\label{eq:bw-weyl}
}
where $k_{\pm}=k_1\pm i k_2$ and $\sigma_{\pm}=\frac{\sigma_1\pm i \sigma_2}{2}$.
The Weyl points appear at
\begin{align}
k_{3}^{W,\pm}=\frac{(2\mp1)\Omega+\sqrt{((2\pm1)\Omega)^{2}-8A^{2}}}{4},\; k_{1,2}=0,
\label{eq:k3w}
\end{align}
which agree with the numerical results as shown in Fig.~\ref{fig:comp}. Eq.~(\ref{eq:bw-weyl}) further implies that the monopole charges are  $+1$ and $+3$. One can generalize this discussion to $k_3\sim{n}\Omega$ with $\pm{n}$-photon sectors, where the off diagonal term is proportional to $A^n (k_+^{2n\mp1}\sigma^++\text{h.c.})$ to leading orders in $A$ implying monopole numbers of $2n\mp1$. 
In the negative $k_3<0$ side, we obtain Weyl nodes with $-2n\pm 1$ monopole numbers, and the pair annihilation shown in Fig.~\ref{fig:annihilation} occurs between those with opposite monopole numbers. 
We note that similar nodal states with nontrivial winding were studied in the two dimensional problem as well \cite{Sentef:2016NC}. 

\paragraph{Emergent Fermi arcs and lattice effects:}
\begin{figure}[t]
\includegraphics[width=\columnwidth]{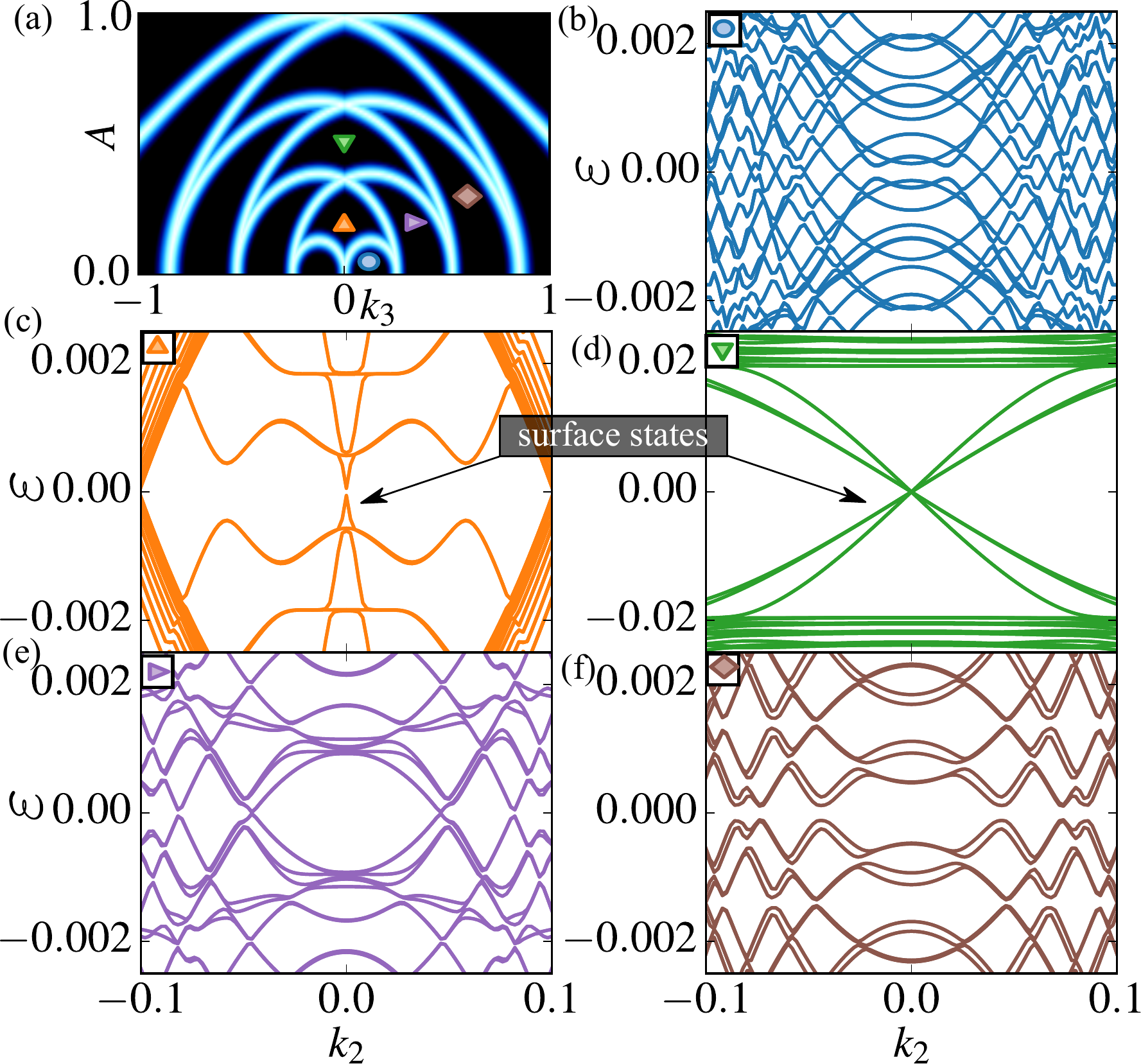}
\caption{
(a) Weyl node trajectories similar to Fig.~\ref{fig:weyltraj} for the lattice system with $\Omega=0.25$. The markers in the plot correspond to the values of $A$ and $k_3$ for which the quasienergy spectrum with open boundary conditions is shown. In plots (b), (e) and (f) there are no surface states; in (c) and (d) there are respectively one  and three surface states. From the change in the number of Fermi arcs between the regions marked by the blue circle and the orange and purple triangles, it can be deduced that the monopole charges of the Weyl nodes from the $n=\pm1$ sectors are $1$ and $-1$ respectively. Similarly, from the regions marked with the green and purple triangles and the brown diamond, it can be deduced that the monopole charges of the Weyl nodes $n=\pm2$ sectors are $\pm3$. Note that, each of the Fermi arcs shown here are two-fold degenerate due to equivalent contributions from $k_1=0$ and $\pi$. The system has a linear dimension of 512 and $\mcl{H}^F$ is truncated till $n=4$.}
\label{fig:arcs}
\end{figure}
In a system with open boundary conditions, the monopole charge of the Weyl nodes can then be deduced from the change in the number and chirality of the Fermi arcs edge states across the Weyl nodes. 
Equivalently, a WSM can be viewed as a momentum-space stack (along $k_3$ in this case) of two-dimensional Chern insulators with a $k_3$ dependent mass, where the Weyl nodes serve as points of topological phase transitions leading to a change in the Chern number and hence in the number of edge states. 
In order to reconcile the number of edge states with the monopole charges of the Floquet Weyl nodes, we regularize our continuum theory on a four-orbital cubic lattice and study the quasienergy spectrum of with open boundary conditions along $x$ but periodic along $y$ and $z$.
\begin{figure}[t]
\includegraphics[width=\columnwidth]{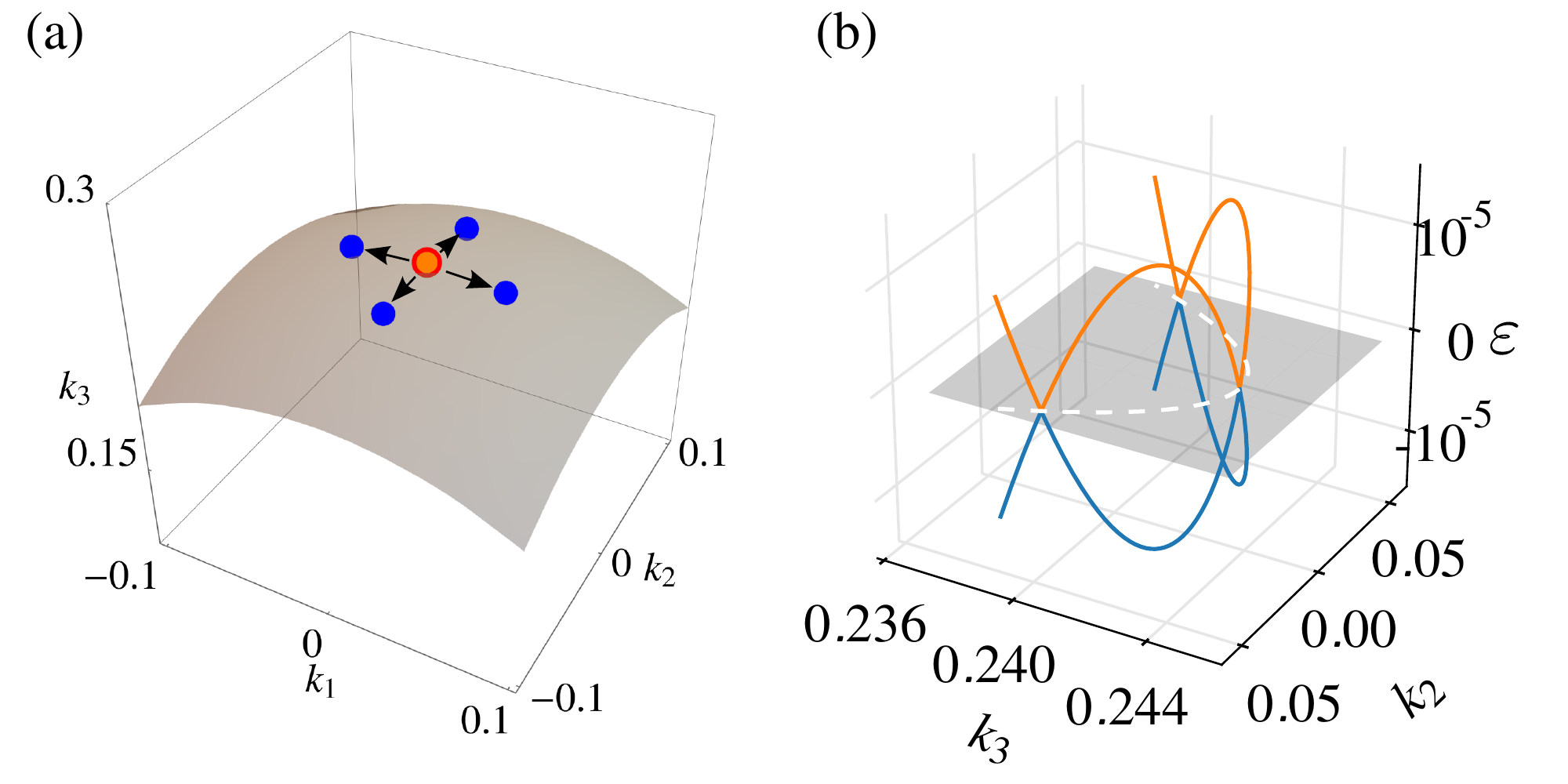}
\caption{
Splitting of Weyl points due to $C{-}4$ symmetric lattice. (a) The original Weyl point obtained from the continuum theory with monopole charge of $+3$ (red) is split into four $C{-}4$ symmetric Weyl points (blue) with charges of $+1$ each while the monopole charge of the original Weyl point on the $k_3$-axis now becomes -1 (orange). (b) The quasienergy bands along the white dashed line are shown, which corresponds to the minima of the quasienergy spectrum in the $k_2$-$k_3$ plane for $k_1=0$ illustrating the additional Weyl points apart from the one at $k_2=0$. The parameters used are $\Omega=0.25$ and $A=0.08$.
}
\label{fig:fringe}
\end{figure}
Here, using a four orbital fermion operator $\Psi_{\hat{\mu}}$, we consider a lattice model 
\eq{
\mcl{H}(t)=-\frac{1}{2}\sum_\pr\sum_{\mu=1}^3(\imath\Psi_\pr^\dagger\gamma^0\gamma^\mu\Psi_{\pr+\hat{\mu}}e^{-\imath\mathbf{A}(t)\cdot\hat{\mu}}+\mrm{h.c.}).
\label{eq:htlatt}
}
as a regularized version of the DSM in CPL incorporated as a time-dependent Peierl's substitution. 
In momentum space, it reduces to $\mcl{H}_{\mrm{DSM}}{=}{-}\sum_\vk \sum_{\mu=1}^3\gamma^0\gamma^\mu\sin (k_\mu-A_\mu)$. 
Using Eq.~\eqref{eq:hmn}, the components of the Floquet Hamiltonian for the lattice system becomes
\eq{
\mcl{H}^F_m &= -\frac{1}{2}J_m(A)\left[ (-\imath)^m\lambda_1 +(-\imath)^{-m}\lambda_1^\dagger +\lambda_2 +(-1)^m\lambda_2^\dagger\right]\nn\\  
\mcl{H}^F_0 &= -\frac{1}{2}\left[J_0(A)(\lambda_1+\lambda_2) + \lambda_3\right] + \mrm{h.c.},
}
where $\lambda_\mu = \imath\sum_\pr\Psi_\pr^\dagger\gamma^0\gamma^\mu\Psi_{\pr+\hat{\mu}}$ and $J_m$ is the $m^{\mrm{th}}$ Bessel function of the first kind.

The lattice Hamiltonian allows direct transitions between all photon sectors
and may result in a difference in the monopole charge of Weyl nodes compared with the continuum theory. 
For instance, $\mathcal{H}^F_{\pm2}$ give an additional off-diagonal term $\sim-(A^{2}/8)k_{-}\sigma^++\text{h.c.}$ to $\mathcal{H}_\mrm{eff}^{W,-}$ in Eq.~(\ref{eq:bw-weyl}). This term changes the monopole charge from $+3$ to $-1$ in the vicinity of $k_3=k_3^{W,-}$. 
This is manifested in the number of the Fermi arcs changing by one across this Weyl point as shown in the regions corresponding to the blue circle  and  the orange and purple triangles in Fig~\ref{fig:arcs}.
Regularizing the theory on a $C-4$ symmetric lattice essentially leads to four additional Weyl points with charges $+1$ branching out away from the $k_3$-axis such that their net charges and that of the one on the $k_3$-axis sum to $-3$ which is shown in Fig.~\ref{fig:fringe}.
A simple counting of edge states in Fig.~\ref{fig:arcs} also shows that that the monopole charges of the Weyl nodes generated from the $n=\pm2$-photon sector are $\pm3$.
%

\paragraph{Conclusions:}
To summarize, we have derived an effective low energy theory for a Floquet WSM obtained by subjecting a DSM to CPL. We especially focused on the Weyl points originating from the Floquet replica in the resonant limit. We found that tuning the frequency or the amplitude of the CPL can move these Weyl points such that they can merge and annihilate in pairs. We also found that such a Floquet WSM allows for Weyl points of higher monopole charges, which we finally reconciled by numerically studying the number of Fermi arcs on a lattice system.

The annihilation process of Weyl nodes is experimentally accessible. The Fermi velocity of the DSM Cd$_3$As$_2$ is $1.5\times10^6$ $\mbox{m/s}$.\cite{neupane2014observation} If a mid infrared laser with photon energy $\Omega=0.2$ $\mbox{eV}$ is used, the pair annihilation taking place at $A/\Omega=0.5$ can be achieved at $E\simeq 0.2 $ $\mbox{MeV/cm}$.  Our results can potentially be verified using time resolved ARPES, as it has already been successfully performed in \cite{wang2013observation,mahmood2016selective} to obtain the Floquet replicas in Bi$_2$Se$_3$. 

\begin{acknowledgments}
\paragraph{Acknowledgments:} 
This work is partially supported by KAKENHI (Grant No. 23740260) and from the ImPact project (No. 2015-PM12-05-01) from JST.
\end{acknowledgments}

\bibliography{refs}

\begin{widetext}
\setcounter{equation}{0}
\setcounter{figure}{0}

\renewcommand{\thefigure}{S\arabic{figure}}
\renewcommand{\theequation}{S\arabic{equation}}

\begin{center}
 \textbf{Supplementary material for ``Emergent Weyl nodes and Fermi arcs in a  Floquet Weyl semimetal''}\\
Leda Bucciantini, Sthitadhi Roy, Sota Kitamura, and Takashi Oka
\end{center}

\section{Brillouin Wigner Expansion}
In this section we will derive the effective Hamiltonian following the Brillouin Wigner Expansion. The Brillouin-Wigner expansion reduces an eigenvalue problem 
\begin{equation}\label{uno}
\sum_{m}(\mathcal{H}_{n,m}^{F}-\delta_{n,m}n\Omega)|\bm{u}_{m}\rangle=\varepsilon|\bm{u}_{n}\rangle
\end{equation}
in the extended Hilbert space to an eigenvalue problem in a smaller Hilbert space, obtained by applying a projection operator $P$ on a certain model space, i.e. 
\begin{equation}
\sum_{m,l}\mathcal{H}_{n,m}^{\text{BW}}P_{m,l}|\bm{u}_{l}\rangle=\varepsilon\sum_{m}P_{n,m}|\bm{u}_{m}\rangle.\label{eq:two}
\end{equation}

Here we are interested in the low energy structure due to the $n=\pm1$ photon sectors, so that we take $P$ as $P_{n,m}=\delta_{n,1}\delta_{m,1}+\delta_{n,-1}\delta_{m,-1}$.

To derive the Brillouin-Wigner Hamiltonian in a series form, let us first divide the original Hamiltonian into two parts $[\mathcal{H}_{0}-\mathcal{N}\Omega]_{n,m}=\delta_{n,m}(\mathcal{H}_{n,n}^{F}-n\Omega)$ and $[\tilde{\mathcal{H}}]_{n,m}=(1-\delta_{n,m})\mathcal{H}_{n,m}^{F}$%
\footnote{This division is suitable for $A$ expansion, while another choice, a division into $-\delta_{n,m}n\Omega$ and $\mathcal{H}_{n,m}^{F}$, is suitable for $\Omega^{-1}$ expansion.%
}. 
By applying $Q_{n,m}=\delta_{n,m}-P_{n,m}$ on Eq. (\ref{uno}), one obtains
\begin{equation}\label{three}
Q|\bm{u}\rangle=Q\dfrac{1}{\varepsilon+\mathcal{N}\Omega-\mathcal{H}_{0}}Q\tilde{\mathcal{H}}|\bm{u}\rangle.
\end{equation}
Here we have suppressed the suffices. This equation reads a recursive equation for an eigenvector $|\bm{u}\rangle$,
\begin{align}\label{four}
|\bm{u}\rangle& = P|\bm{u}\rangle+Q\dfrac{1}{\varepsilon+\mathcal{N}\Omega-\mathcal{H}_{0}}Q\tilde{\mathcal{H}}|\bm{u}\rangle\nonumber \\
 & =P|\bm{u}\rangle+Q\dfrac{1}{\varepsilon+\mathcal{N}\Omega-\mathcal{H}_{0}}Q\tilde{\mathcal{H}}P|\bm{u}\rangle+Q\dfrac{1}{\varepsilon+\mathcal{N}\Omega-\mathcal{H}_{0}}Q\tilde{\mathcal{H}}Q\dfrac{1}{\varepsilon+\mathcal{N}\Omega-\mathcal{H}_{0}}Q\tilde{\mathcal{H}}P|\bm{u}\rangle+\dots.
\end{align}

Then, by acting with $P(\mathcal{H}^{F}-\mathcal{N}\Omega)$ on (\ref{four}) one can extract $P|\bm{u}\rangle$
\begin{equation}
\varepsilon P|\bm{u}\rangle=P(\mathcal{H}^{F}-\mathcal{N}\Omega)P|\bm{u}\rangle+P\tilde{\mathcal{H}}Q\dfrac{1}{\varepsilon+\mathcal{N}\Omega-\mathcal{H}_{0}}Q\tilde{\mathcal{H}}P|\bm{u}\rangle+P\tilde{\mathcal{H}}Q\dfrac{1}{\varepsilon+\mathcal{N}\Omega-\mathcal{H}_{0}}Q\tilde{\mathcal{H}}Q\dfrac{1}{\varepsilon+\mathcal{N}\Omega-\mathcal{H}_{0}}Q\tilde{\mathcal{H}}P|\bm{u}\rangle+\dots,
\end{equation}
which implies, comparing with (\ref{eq:two}), that the effective Hamiltonian $\mathcal{H}^{\text{BW}}$ on the subspace identified by the projector $P$ is
\begin{equation}
\mathcal{H}^{\text{BW}}=P(\mathcal{H}^{F}-\mathcal{N}\Omega)P+P\tilde{\mathcal{H}}Q\dfrac{1}{\varepsilon+\mathcal{N}\Omega-\mathcal{H}_{0}}Q\tilde{\mathcal{H}}P+P\tilde{\mathcal{H}}Q\dfrac{1}{\varepsilon+\mathcal{N}\Omega-\mathcal{H}_{0}}Q\tilde{\mathcal{H}}Q\dfrac{1}{\varepsilon+\mathcal{N}\Omega-\mathcal{H}_{0}}Q\tilde{\mathcal{H}}P+\dots.
\end{equation}

We will then apply it to the present DSM system driven by the CPL. With the chiral representation, the Floquet components can be written as 
\begin{gather}
\mathcal{H}_{n,n}^{F}=\Gamma\begin{pmatrix}k_{3} & k_{1}-ik_{2}\\
k_{1}+ik_{2} & -k_{3}
\end{pmatrix},\,\mathcal{H}_{n+1,n}^{F}=-\Gamma\begin{pmatrix}0 & A\\
0 & 0
\end{pmatrix},\,\mathcal{H}_{n,n+1}^{F}=-\Gamma\begin{pmatrix}0 & 0\\
A & 0
\end{pmatrix},
\end{gather}
and $\mathcal{H}_{m,n}^{F}=0$ for other $m,n$. Here the Hamiltonian
is block-diagonal in $\gamma^{5}$, so that we denote the eigenvalue
of $-\gamma^{5}$ as $\Gamma=\pm1$ and have shown the block Hamiltonian.

One can further transform the basis to diagonalize $\mathcal{H}_{n,n}^{F}$
by, 
\begin{equation}
U=\begin{pmatrix}\cos\dfrac{\theta}{2} & -e^{-i\phi}\sin\dfrac{\theta}{2}\\
e^{i\phi}\sin\dfrac{\theta}{2} & \cos\dfrac{\theta}{2}
\end{pmatrix}\,(\cos\theta>0),\,\begin{pmatrix}-\sin\dfrac{\theta}{2} & -e^{-i\phi}\cos\dfrac{\theta}{2}\\
e^{i\phi}\cos\dfrac{\theta}{2} & -\sin\dfrac{\theta}{2}
\end{pmatrix}\,(\cos\theta<0),
\end{equation}
as
\begin{gather}
U^{\dagger}\mathcal{H}_{n,n}^{F}U=\begin{pmatrix}\Gamma|\bm{k}| & 0\\
0 & -\Gamma|\bm{k}|
\end{pmatrix},\, U^{\dagger}\mathcal{H}_{n\pm1,n}^{F}U=-\dfrac{\Gamma A}{2}\begin{pmatrix}e^{\pm i\phi}\sin\theta & \pm ge^{i(\pm1-g)\phi}(1\pm\cos\theta)\\
\mp ge^{i(\pm1+g)\phi}(1\mp\cos\theta) & -e^{\pm i\phi}\sin\theta
\end{pmatrix},
\end{gather}
where $g=\text{sgn}k_{3}$.For the present choice of $P$, $\mathcal{H}^{\text{BW}}$
can be explicitly written as
\begin{align}
\mathcal{H}^{\text{BW}} & =\begin{pmatrix}\mathcal{H}_{1,1}^{F}-\Omega & 0\\
0 & \mathcal{H}_{-1,-1}^{F}+\Omega
\end{pmatrix}+\dfrac{1}{\varepsilon^{2}-\bm{k}^{2}}\begin{pmatrix}\mathcal{H}_{1,0}^{F}(\varepsilon+\mathcal{H}_{0,0}^{F})\mathcal{H}_{0,1}^{F} & \mathcal{H}_{1,0}^{F}(\varepsilon+\mathcal{H}_{0,0}^{F})\mathcal{H}_{0,-1}^{F}\\
\mathcal{H}_{-1,0}^{F}(\varepsilon+\mathcal{H}_{0,0}^{F})\mathcal{H}_{0,1}^{F} & \mathcal{H}_{-1,0}^{F}(\varepsilon+\mathcal{H}_{0,0}^{F})\mathcal{H}_{0,-1}^{F}
\end{pmatrix}\nonumber \\
 & +\dfrac{1}{(\varepsilon+2\Omega)^{2}-\bm{k}^{2}}\begin{pmatrix}\mathcal{H}_{1,2}^{F}(\varepsilon+2\Omega+\mathcal{H}_{2,2}^{F})\mathcal{H}_{2,1}^{F} & 0\\
0 & 0
\end{pmatrix}\nonumber \\
 & +\dfrac{1}{(\varepsilon-2\Omega)^{2}-\bm{k}^{2}}\begin{pmatrix}0 & 0\\
0 & \mathcal{H}_{-1,-2}^{F}(\varepsilon-2\Omega+\mathcal{H}_{-2,-2}^{F})\mathcal{H}_{-2,-1}^{F}
\end{pmatrix}+\mathcal{O}(A^{3}).
\end{align}
Here we have used $\mathcal{H}_{n,n}^{2}=\bm{k}^{2}$. With the basis
diagonal in $\mathcal{H}_{n,n}^{F}$, we obtain 
\begin{align}
\mathcal{H}^{\text{BW}} & =\begin{pmatrix}\Gamma|\bm{k}|-\Omega & 0 & 0 & 0\\
0 & -\Gamma|\bm{k}|-\Omega & 0 & 0\\
0 & 0 & \Gamma|\bm{k}|+\Omega & 0\\
0 & 0 & 0 & -\Gamma|\bm{k}|+\Omega
\end{pmatrix}\nonumber \\
 & +\dfrac{A^{2}}{2}\dfrac{\varepsilon-\Gamma k_{3}}{\varepsilon^{2}-\bm{k}^{2}}\begin{pmatrix}1+\cos\theta & -ge^{-ig\phi}\sin\theta & 0 & 0\\
-ge^{ig\phi}\sin\theta & 1-\cos\theta & 0 & 0\\
0 & 0 & 0 & 0\\
0 & 0 & 0 & 0
\end{pmatrix}+\dfrac{A^{2}}{2}\dfrac{\varepsilon+\Gamma k_{3}}{\varepsilon^{2}-\bm{k}^{2}}\begin{pmatrix}0 & 0 & 0 & 0\\
0 & 0 & 0 & 0\\
0 & 0 & 1-\cos\theta & ge^{-ig\phi}\sin\theta\\
0 & 0 & ge^{ig\phi}\sin\theta & 1+\cos\theta
\end{pmatrix}\nonumber \\
 & +\dfrac{A^{2}}{2}\dfrac{\varepsilon+2\Omega+\Gamma k_{3}}{(\varepsilon+2\Omega)^{2}-\bm{k}^{2}}\begin{pmatrix}1-\cos\theta & ge^{-ig\phi}\sin\theta & 0 & 0\\
ge^{ig\phi}\sin\theta & 1+\cos\theta & 0 & 0\\
0 & 0 & 0 & 0\\
0 & 0 & 0 & 0
\end{pmatrix}+\dfrac{A^{2}}{2}\dfrac{\varepsilon-2\Omega-\Gamma k_{3}}{(\varepsilon-2\Omega)^{2}-\bm{k}^{2}}\begin{pmatrix}0 & 0 & 0 & 0\\
0 & 0 & 0 & 0\\
0 & 0 & 1+\cos\theta & -ge^{-ig\phi}\sin\theta\\
0 & 0 & -ge^{ig\phi}\sin\theta & 1-\cos\theta
\end{pmatrix}\nonumber \\
 & +\dfrac{A^{2}}{2}\dfrac{\Gamma|\bm{k}|\sin\theta}{\varepsilon^{2}-\bm{k}^{2}}\begin{pmatrix}0 & 0 & e^{i2\phi}\sin\theta & ge^{i(2-g)\phi}(1+\cos\theta)\\
0 & 0 & -ge^{i(2+g)\phi}(1-\cos\theta) & -e^{i2\phi}\sin\theta\\
e^{-i2\phi}\sin\theta & -ge^{-i(2+g)\phi}(1-\cos\theta) & 0 & 0\\
ge^{-i(2-g)\phi}(1+\cos\theta) & -e^{-i2\phi}\sin\theta & 0 & 0
\end{pmatrix}+\mathcal{O}(A^{3}).
\end{align}
One can further project out high-energy states with $\varepsilon=\pm(|\bm{k}|+\Omega)+\mathcal{O}(A^{2})$
by repeating the same procedure. Then we obtain two Hamiltonians for
$\Gamma=\pm1$ as 
\begin{align}
\mathcal{H}_{\text{eff}}^{\Gamma=\pm1} & =\begin{pmatrix}|\bm{k}|-\Omega & 0\\
0 & -|\bm{k}|+\Omega
\end{pmatrix}\nonumber \\
 & +\dfrac{A^{2}}{2}\dfrac{\varepsilon-\Gamma k_{3}}{\varepsilon^{2}-\bm{k}^{2}}\begin{pmatrix}1+\Gamma\cos\theta & 0\\
0 & 0
\end{pmatrix}+\dfrac{A^{2}}{2}\dfrac{\varepsilon+\Gamma k_{3}}{\varepsilon^{2}-\bm{k}^{2}}\begin{pmatrix}0 & 0\\
0 & 1+\Gamma\cos\theta
\end{pmatrix}\nonumber \\
 & +\dfrac{A^{2}}{2}\dfrac{\varepsilon+2\Omega+\Gamma k_{3}}{(\varepsilon+2\Omega)^{2}-\bm{k}^{2}}\begin{pmatrix}1-\Gamma\cos\theta & 0\\
0 & 0
\end{pmatrix}+\dfrac{A^{2}}{2}\dfrac{\varepsilon-2\Omega-\Gamma k_{3}}{(\varepsilon-2\Omega)^{2}-\bm{k}^{2}}\begin{pmatrix}0 & 0\\
0 & 1-\Gamma\cos\theta
\end{pmatrix}\nonumber \\
 & +\dfrac{A^{2}}{2}\dfrac{gk\sin\theta(1+\Gamma\cos\theta)}{\varepsilon^{2}-\bm{k}^{2}}\begin{pmatrix}0 & e^{i(2-\Gamma g)\phi}\\
e^{-i(2-\Gamma g)\phi} & 0
\end{pmatrix}+\mathcal{O}(A^{3}),
\end{align}
and by using $\varepsilon^{2}=(|\bm{k}|-\Omega)^{2}+\mathcal{O}(A^{2})$
and $\varepsilon\mathds{1}_{2}=\mathcal{H}_{\text{eff}}$, finally one
obtains 
\begin{equation}
\mathcal{H}_{\text{eff}}^{\Gamma=\pm1}=\begin{pmatrix}|\bm{k}|-\Omega & 0\\
0 & -|\bm{k}|+\Omega
\end{pmatrix}+A^{2}\dfrac{|\bm{k}|^{2}+k_{3}^{2}+\Gamma\Omega k_{3}}{|\bm{k}|(4|\bm{k}|^{2}-\Omega^{2})}\begin{pmatrix}1 & 0\\
0 & -1
\end{pmatrix}-gA^{2}\dfrac{(|\bm{k}|+|k_{3}|)^{\Gamma g}}{2|\bm{k}|\Omega(2|\bm{k}|-\Omega)}\begin{pmatrix}0 & k_{+}^{2-\Gamma g}\\
k_{-}^{2-\Gamma g} & 0
\end{pmatrix}+\mathcal{O}(A^{3}).
\end{equation}

By imposing that the vanishing of the $\sigma_i$ coefficients in the two previous Hamiltonians and retaining the meaningful solutions, we obtain the positions of the Weyl points as in Eq.(8) of the main text.
The expanded form around the Weyl point is 
\begin{align}
\mathcal{H}_{\text{eff}}^{\Gamma=g} & \sim\left[|k_{3}|-\Omega+\dfrac{A^{2}}{\Omega}\right]\begin{pmatrix}1 & 0\\
0 & -1
\end{pmatrix}-g\dfrac{A^{2}}{\Omega^{2}}(k_{1}^{2}+k_{2}^{2})^{1/2}\begin{pmatrix}0 & e^{i\phi}\\
e^{-i\phi} & 0
\end{pmatrix},\\
\mathcal{H}_{\text{eff}}^{\Gamma=-g} & \sim\left[|k_{3}|-\Omega+\dfrac{A^{2}}{3\Omega}\right]\begin{pmatrix}1 & 0\\
0 & -1
\end{pmatrix}-g\dfrac{A^{2}}{4\Omega^{4}}(k_{1}^{2}+k_{2}^{2})^{3/2}\begin{pmatrix}0 & e^{i3\phi}\\
e^{-i3\phi} & 0
\end{pmatrix}.
\end{align}
One can calculate the monopole charge from the Chern number for fixed $k_{3}$ and its jump at the Weyl point; it yields $C=2g-\Gamma=+g,+3g$.

\section{Corrections to effective low energy theory for a lattice}
In this section we discuss how the regularization on a cubic lattice affects the Weyl points emerging near $k_{3}\simeq\pm\Omega$. The regularized Hamiltonian
\[
H(t)=\Gamma\begin{pmatrix}\sin k_{3} & \sin(k_{1}-A\cos\Omega t)-i\sin(k_{2}-A\sin\Omega t)\\
c.c. & -\sin k_{3}
\end{pmatrix},
\]
coincides with the continuous model if $A,k,\Omega\ll1$. To include the correction to the continuous model, here we expand the lattice Hamiltonian up to second order in $A$ and third order in $\bm{k}$;
\begin{align}
H(t) & =\Gamma\vert\vk\vert\begin{pmatrix}\cos\theta & \sin\theta e^{-i\phi}\\
c.c. & -\cos\theta
\end{pmatrix}-\dfrac{1}{6}\Gamma\vert\vk\vert^{3}\begin{pmatrix}\cos^{3}\theta & \sin^{3}\theta(\cos^{3}\phi-i\sin^{3}\phi)\\
c.c. & -\cos^{3}\theta
\end{pmatrix}\nonumber \\
 & -\Gamma\dfrac{A^{2}}{4}\vert\vk\vert\sin\theta\begin{pmatrix}0 & e^{-i\phi}\\
c.c. & 0
\end{pmatrix}+\Gamma\dfrac{A^{2}}{4}\dfrac{\vert\vk\vert^{3}}{6}\sin^{3}\theta\begin{pmatrix}0 & (\cos^{3}\phi-i\sin^{3}\phi)\\
c.c. & 0
\end{pmatrix}\nonumber \\
 & -\Gamma A\begin{pmatrix}0 & -\dfrac{\vk^{2}\sin^{2}\theta\cos2\phi}{4}\\
1-\dfrac{\vk^{2}\sin^{2}\theta}{4} & 0
\end{pmatrix}e^{i\Omega t}-\Gamma A\begin{pmatrix}0 & 1-\dfrac{\vk^{2}\sin^{2}\theta}{4}\\
-\dfrac{\vk^{2}\sin^{2}\theta\cos2\phi}{4} & 0
\end{pmatrix}e^{-i\Omega t}\nonumber \\
 & -\Gamma\dfrac{A^{2}}{8}\vert\vk\vert\sin\theta\begin{pmatrix}0 & e^{i\phi}\\
c.c. & 0
\end{pmatrix}(e^{i2\Omega t}+e^{-i2\Omega t})+\Gamma\dfrac{A^{2}}{8}\dfrac{\vert\vk\vert^{3}}{6}\sin^{3}\theta\begin{pmatrix}0 & \cos^{3}\phi+i\sin^{3}\phi\\
c.c. & 0
\end{pmatrix}(e^{i2\Omega t}+e^{-i2\Omega t}).
\end{align}

Deriving the Brillouin-Wigner Hamiltonian up to order $A^{2}$ and $\Omega$ yields
\begin{align}\label{eq11}
\mathcal{H}_{\text{eff}} & =(R-\Omega)\sigma_{3}-A^{2}\left(\dfrac{\vert\vk\vert\sin^{2}\theta-(2R+\Omega\Gamma\cos\theta)}{4R^{2}-\Omega^{2}}\right)\sigma_{3}\nonumber \\
 & -A^{2}\left(\dfrac{(2R+\Omega\Gamma\cos\theta)\vk^{2}\sin^{2}\theta}{2(4R^{2}-\Omega^{2})}+\dfrac{\vert\vk\vert}{4}\sin^{2}\theta-\dfrac{7(k_{1}^{4}+k_{2}^{4})}{6k(4R^{2}-\Omega^{2})}\right)\sigma_{3}\nonumber \\
 & +gA^{2}k\sin\theta\left(-\dfrac{(1+\Gamma\cos\theta)}{2\Omega(2R-\Omega)}-\dfrac{1+\Gamma\cos\theta}{16}+\dfrac{\vk^{2}\sin^{2}\theta}{16\Omega(2R-\Omega)}(5+7\Gamma\cos\theta)\right)e^{i(2-\Gamma g)\phi}\sigma_{+}+h.c.\nonumber \\
 & +gA^{2}k\sin\theta\left(\dfrac{1-\Gamma\cos\theta}{16}+\dfrac{\vk^{2}\sin^{2}\theta}{48\Omega(2R-\Omega)}(1+7\Gamma\cos\theta)\right)e^{-i(2+\Gamma g)\phi}\sigma_{+}+h.c.
\end{align}
with 
\begin{equation}
R=|\bm{k}|-\dfrac{1}{6|\bm{k}|}\sum_{i}k_{i}^{4}.
\end{equation}
We can confirm that the continuous model can be recovered by taking $\Omega\sim k\rightarrow0$. After imposing that the $\sigma_{i}$ coefficients vanish, we obtain that the Weyl points appear not only along the $k_{3}$ axis but also lie in the $(\theta,\phi)$ plane.

The Hamiltonian around the Weyl points lying on the $k_{3}$ axis
is expressed as 
\begin{equation}
\mathcal{H}_{\text{eff}}^{\Gamma=+g}=\left[|k_{3}|-\Omega\left(1+\dfrac{\Omega^{2}}{6}\right)+\dfrac{A^{2}}{\Omega}\right]\sigma_{3}-\left[gA^{2}\vert\vk\vert\sin\theta\left(\dfrac{1}{\Omega^{2}}+\dfrac{1}{8}\right)e^{i\phi}\sigma_{+}+h.c.\right],
\end{equation}
\begin{equation}
\mathcal{H}_{\text{eff}}^{\Gamma=-g}=\left[|k_{3}|-\Omega\left(1+\dfrac{\Omega^{2}}{6}\right)+\dfrac{A^{2}}{3\Omega}\right]\sigma_{3}+\left[\dfrac{g}{8}A^{2}\vert\vk\vert\sin\theta e^{-i\phi}\sigma_{+}+h.c.\right],
\end{equation}
where the terms $\propto\sin^{3}\theta$ have been neglected since we are interested in the vicinity of $\theta=0,\pi$. Note that then the offdiagonal term in $\Gamma=-g$
case is purely derived from the nonlinear correction, thus the Chern number is different from that
of continuous model: indeed the relevant term comes from the direct transition via $\mathcal{H}_{\pm2}^{F}$,
which is absent in the continuous model. 
The monopole charges are $C=\Gamma =+g,-g$, different from the results for the continuous model $C=2g-\Gamma =+g,+3g$.

The position of the Weyl points in the $(\theta,\phi)$ plane is obtained by solving the following equations
\begin{align}
R&=\Omega+A^{2}\left(\dfrac{\vert\vk\vert\sin^{2}\theta}{3\Omega^{2}}-\dfrac{2+\Gamma\cos\theta}{3\Omega}+\dfrac{\vk^{2}\sin^{2}\theta}{6\Omega}(2+\Gamma\cos\theta)+\dfrac{|\bm{k}|\sin^{2}\theta}{4}-\dfrac{7(k_{1}^{4}+k_{2}^{4})}{18\vert\vk\vert\Omega^{2}}\right)+\mathcal{O}(A^{3}),\\
0 & =\left(-\dfrac{1+\Gamma\cos\theta}{2\Omega^{2}}-\dfrac{1+\Gamma\cos\theta}{16}+\dfrac{\sin^{2}\theta}{16}(5+7\Gamma\cos\theta)\right)+\left(\dfrac{1-\Gamma\cos\theta}{16}+\dfrac{\sin^{2}\theta}{48}(1+7\Gamma\cos\theta)\right)\cos4\phi,\\
0 & =\left(\dfrac{1-\Gamma\cos\theta}{16}+\dfrac{\sin^{2}\theta}{48}(1+7\Gamma\cos\theta)\right)\sin4\phi.
\end{align}
This entails that $\phi=0,\,\pi/2,\,\pi,\,3\pi/2$ thus implying a
$C-4$ symmetry of the emerging Weyl points and 
\begin{equation}
\dfrac{1}{\Omega^{2}}=\dfrac{8+11\Gamma\cos\theta-8\cos^{2}\theta-14\Gamma\cos^{3}\theta}{12(1+\Gamma\cos\theta)}.\label{eq:costheta}
\end{equation}
While Eq.~(\ref{eq:costheta}) has multiple solutions in the large $\varepsilon$ region, it is an artifact due to the truncation of the nonlinear correction.
Fig.~\ref{fig:traj} shows the trajectory of the Weyl points on the $k_1$-$k_3$ plane when $\Omega$ is varied, calculated with higher order corrections.

It is natural to expect that these four Weyl points have the same monopole charge, 
and then it is consistent with the continuous limit $\Omega\rightarrow0$ if they have $C=+g$: 
In this limit Eq.~(\ref{eq:costheta}) has a solution $\cos\theta\rightarrow-\Gamma$ if $\Gamma=-g$, thus these Weyl points marge with that lying on the $k_{3}$ axis (with $C=\Gamma=-g$). The merged Weyl point has a monopole charge $(+g)\times4+(-g)=+3g$, as predicted in the continuous model.

\begin{figure}
\centering
\includegraphics[width=0.5\textwidth]{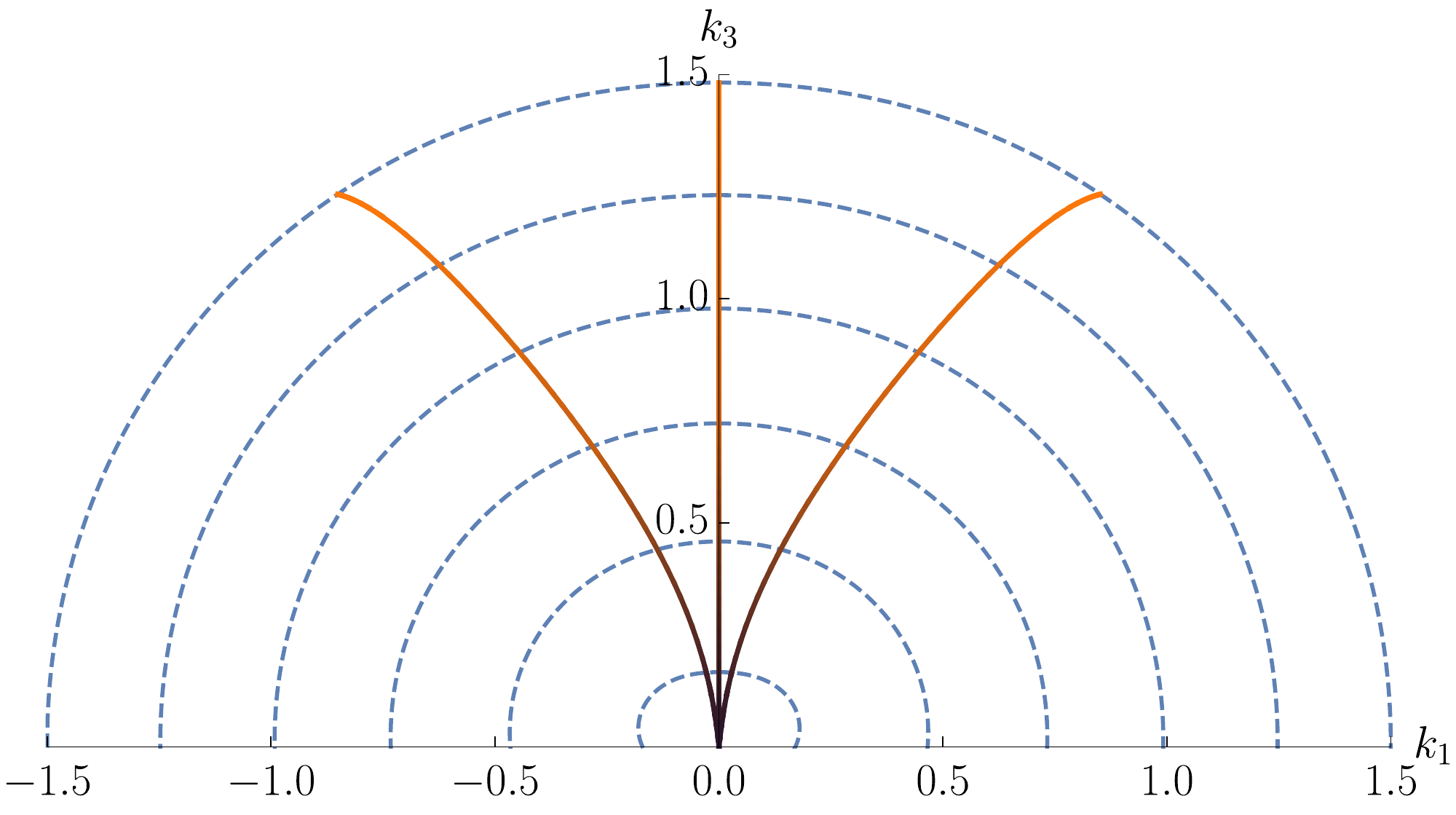}
\caption{\label{fig:traj}The trajectory of the location of Weyl points on the $k_1$-$k_3$ plane when $\Omega$ is varied.
The solid lines represent the Weyl points, while dashed lines show energy minima along radial direction for $\Omega=0.25,0.5,\dots,1.5$.
}
\end{figure}

\end{widetext}
\end{document}